\documentclass[conference]{IEEEtran}

\usepackage{cite}
\usepackage{graphicx}
\usepackage{amssymb,amsmath}
\usepackage{siunitx}
\usepackage{multirow}
\usepackage{xcolor}
\definecolor{redcolor}{rgb}{1, 0, 0}
\usepackage{subfiles}
\usepackage{makecell}
\usepackage{color,soul}
\usepackage{multirow}
\usepackage{mathtools}
\usepackage{csquotes}
\usepackage{esvect}
\usepackage{stfloats}   
\usepackage{float}

\usepackage{caption}
\usepackage{subcaption}

\usepackage{csquotes}
\MakeOuterQuote{"}

\begin{document}

\title{Deep Neuromorphic Networks with Superconducting Single Flux Quanta}


\author{
    \bigskip
	\IEEEauthorblockN{Gleb Krylov$^1$, Alexander J. Edwards$^2$, Joseph S. Friedman$^2$, and Eby G. Friedman$^1$} \\
	\IEEEauthorblockA{$^1$University of Rochester, Rochester, New York, USA \\ $^2$The University of Texas at Dallas, Richardson, Texas, USA}
}

\maketitle

\begin{abstract}
Conventional semiconductor-based integrated circuits are gradually approaching fundamental scaling limits. Many prospective solutions have recently emerged to supplement or replace both the technology on which basic devices are built and the architecture of data processing.
Neuromorphic circuits are a promising approach to computing where techniques used by the brain to achieve high efficiency are exploited. 
Many existing neuromorphic circuits rely on unconventional and useful properties of novel technologies to better mimic the operation of the brain.
One such technology is single flux quantum (SFQ) logic -- a cryogenic superconductive technology in which the data are represented by quanta of magnetic flux (fluxons) produced and processed by Josephson junctions embedded within inductive loops.
The movement of a fluxon within a circuit produces a quantized voltage pulse (SFQ pulse), resembling a neuronal spiking event. These circuits routinely operate at clock frequencies of tens to hundreds of gigahertz, making SFQ a natural technology for processing high frequency pulse trains.

The similarities between SFQ and neuronal spiking has previously been observed; however, prior proposals for SFQ neural networks often require energy-expensive fluxon conversions, involve heterogeneous technologies, or exclusively focus on device level behavior. In this paper, a design methodology for deep single flux quantum neuromorphic networks is presented.
Synaptic and neuronal circuits based on SFQ technology are presented and characterized.
Based on these primitives, a deep neuromorphic XOR network is evaluated as a case study, both at the architectural and circuit levels, achieving wide classification margins.
The proposed methodology does not employ unconventional superconductive devices or semiconductor transistors. The resulting networks are tunable by an external current, making this proposed system an effective approach for scalable cryogenic neuromorphic computing.

\end{abstract}


\section{Introduction}

\begin{figure*}
    \centering
    \includegraphics[width=\textwidth]{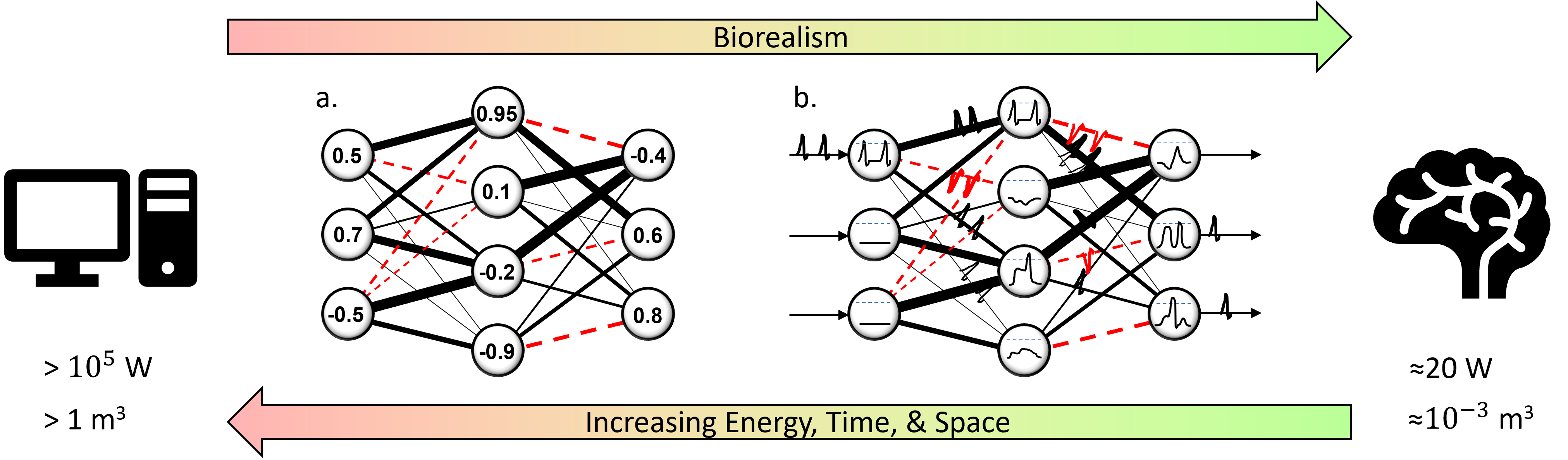}
    \caption{Biological inspiration for spiking neuromorphic networks with SFQ. \textbf{a.} Conventional artificial neural network.  Information propagates through the network from left to right. The neuron states (circles) are encoded by a single real-valued number which is modulated by downstream synapses (black and red lines) which are collected by the successive (post-synaptic) neurons according to an activation function. \textbf{b.} Spiking neural network (SNN).  Rather than representing information as a single number, spiking neurons mimic biology by retransmitting spikes downstream, which are modulated by the synapse strength. The information representation can therefore be much richer, enabling systems approaching the incredible efficiency observed in biology.}
    \label{fig:outlook}
\end{figure*}

Conventional level-based artificial neural networks -- while useful in a large number of applications -- suffer from high computational costs that may be mitigated by using alternative biomimetic architectures such as spiking neuromorphic networks (SNNs) \cite{Schuman2022}.  As the human brain can seemingly compute similar computation at a fraction of the energy, recent attention has gravitated towards biomimetic hardware, emulating biology in both phenomenological behavior and emergent computation.  Mimicking biological neuronal spiking behavior, SNNs are a class of neural networks in which data are encoded in a sequence of temporal spikes as opposed to a single real-valued signal (illustrated in Fig. \ref{fig:outlook}); SNNs are therefore highly attractive for low power neuromorphic hardware.  

Operating on minimal voltage pulses that function like neuronal spiking events, single flux quantum (SFQ) systems are particularly attractive for developing biomimetic SNNs. Superconducting Josephson junctions (JJs) inherently react to and regenerate fluxons, enabling extremely energy efficient neuronal and synaptic circuits.  Whereas modern large scale SNNs utilize inefficient packet-based routing networks \cite{SpiNNaker, TrueNorth, Loihi}, SFQ may be ideally suited for ultra-fast, ultra-low power SNN systems.  While the similarities between SFQ pulses and neuronal spiking have been explored in the literature \cite{goteti21, 6355631, PhysRevE.82.011914, SchneiderSynapses, doi:10.1126/sciadv.1701329, SchneiderFan, PhysRevE.95.032220, SEGALL201471, Schneider_Tutorial, bozbey2020single, BozbeyJJSoma}, a multilayer fully-SFQ neuromorphic network has yet to be demonstrated.


We propose the first multi-layer fully-SFQ neuromorphic network, enabling ultra-low-power neuromorphic computation.  The SFQ network comprises neuronal and synaptic SFQ primitives that may be cascaded with minimal layer-to-layer circuit design to construct several SNN architectures.  SFQ synapses emulating the stochasticity in the brain \cite{MaassSpikeOrNot, BraunStochasticity} are described along with SFQ leaky-integrate-and-fire (LIF) neurons. These primitive synaptic and neuronal circuits directly cascade with each other permitting the construction of deep neuromorphic networks.  A multilayer network computing the XOR functionality is constructed and simulated showing high non-linearity and input separation.

\section{Background}
SNNs and SFQ circuits have received significant attention in the literature including proposals for SNN primitives implemented with SFQ, as we summarize in the following subsections.




\subsection{Spiking Neuromorphic Networks}


SNNs attempt to replicate biological spiking behavior so as to unlock the ultra-low-power computation observed in biological systems.  Biological neural networks encode information based on the frequency and relative timing of neuronal spiking events.  This behavior is counter to conventional software neural networks, which represent information as single-valued numbers.  Spiking neuromorphic networks are an attempt to more richly encode information with equivalent or cheaper hardware for more efficient, biomimetic computation.

Neurons in SNNs mimic biology by aggregating spiking activity from upstream neurons and firing when sufficiently stimulated, creating a spike event that is propagated downstream.  A common neuronal model, the leaky-integrate-and-fire (LIF) neuron, integrates input spikes into an internal potential which leaks over time if not stimulated.  When the potential crosses the action potential threshold, the neuron fires, resets, and begins integrating again.  The LIF model is a very common neuronal model for SNNs \cite{Schuman2022}.

Synapses in SNNs mimic biology by modulating the strengths of the connections between upstream and downstream neurons.  Most SNNs employ synapses that modulate the \textit{amplitude} of spiking events, whereas alternative approaches may modulate spike \textit{rate} or  -- inspired by the prevalent stochasticity observed in biological systems \cite{MaassSpikeOrNot, BraunStochasticity} -- stochastically gate individual synaptic spiking events, the method used here.


\label{sec:src_training}
Spike rate coded spiking neural networks may be trained similar to level-based feed-forward artificial neural networks (ANNs) via backpropagation, assuming the following:

\begin{itemize}
    \item information is only coded in the average spike rate and no useful information is present in the spike timing,
    \item the synapses modulate the spike rate by a trainable constant multiplier, and
    \item the neuronal output spike rate depends only on the input spike rate according to a continuous activation function.
\end{itemize}

Under these assumptions, the output rate of a synapse randomly passing a fixed proportion of fluxons is equal to multiplying the input rate by that synapse weight. Furthermore, confluence of post-synaptic spike trains result in a single spike train whose rate is the sum of the input rates.  This rate may saturate at large frequencies which may be treated like saturation of an activation function. There therefore exists a one-to-one mapping between feed-forward spike-rate-coded SNNs and level-based ANNs, enabling identical training via backpropagation for both networks.  

The benefits of this equivalence are manifold: as spiking SFQ networks may be lower cost than level-based ANN, previously trained networks may be based on SFQ hardware, and new networks may be effectively trained with extant software tools.  Furthermore, training via backpropagation under spike-rate-coding assumptions may generate starting point networks for advanced learning techniques that can enhance the network to encode information in spike timing, further approaching the richness of information coding in the brain.

\subsection{Single Flux Quanta Circuitry}

Single flux quantum logic is an emerging  cryogenic technology for highly energy efficient computing \cite{krylov2022book}. SFQ circuits are based on Josephson junctions (JJs) and superconducting quantum interference devices (SQUIDs), and operate with magnetic flux quanta. The quanta are typically represented by voltage pulses of quantized area equal to the magnetic flux quantum ($\Phi_0$ $\sim$~2.07~mV $\cdot$ ps) \cite{likharev1991rsfq}. These pulses are generated in a process often referred to as JJ switching -- a shift of superconducting phase between the terminals of the JJ by $2\pi$.

The primary advantage of SFQ circuits for digital logic is the unparalleled energy efficiency. Each $2\pi$ transition of a typical 100 $\mu$A JJ dissipates energy on the order of \num{2e-19}~J \cite{mukhanov2011energy}. Although a logic operation requires several switches, the energy per operation is several orders of magnitude lower than state-of-the-art CMOS logic even accounting for the cryogenic cooling to 4.2~K (liquid helium temperature) \cite{holmes2013energy}. 

Whereas in conventional SFQ logic, information is encoded as the presence or absence of an SFQ pulse within a specific time period, alternative JJ-based circuits can generate and operate on more complex SFQ pulse sequences.  This includes the generation of pulse sequences with a controllable frequency and stochastic switching induced by thermal noise. Both properties are exploited in this work.

Several approaches for neuromorphic computing with SFQ circuits have been proposed, however a fully-SFQ multilayer neuromorphic network remains to be demonstrated.  Several proposals demonstrate individual SFQ gates for neuronal \cite{goteti21, 6355631, PhysRevE.82.011914, BozbeyJJSoma}, synaptic \cite{PhysRevE.82.011914, SchneiderSynapses, doi:10.1126/sciadv.1701329, SegallSTDPSynapse_2023}, or interconnect \cite{SchneiderFan} functionality.  Two-neuron oscillatory Hopfield networks \cite{PhysRevE.95.032220, SEGALL201471} and single layer feed-forward Spiking networks \cite{Schneider_Tutorial, schneider_review_2022} have been proposed, and a multi-layer feed-forward spiking network with heterogeneous CMOS-SFQ circuitry has been demonstrated in simulation \cite{bozbey2020single}.
Some of the proposed approaches utilize magnetic Josephson junctions to gradually modify the internal state of the gates, enabling online learning \cite{SchneiderSynapses}, \cite{Schneider_Tutorial}. Fabrication processes used to manufacture these devices are, however, not well established, and the resulting circuits are limited in scale.
In \cite{semenov2022biosfq}, an SFQ-based methodology for building neuromorphic networks is proposed, where a bipolar current is used to represent a logic state, which is converted into a train of SFQ pulses for transmission.

\section{Deep Neuromorphic Networks with SFQ Primitive Circuits}

\label{sec:3}

The first fully SFQ deep neuromorphic network is described here; primitive SFQ synaptic and neuronal circuits digitally manipulating SFQ spike sequences are described. The stochastic synapse is based on a Josephson balanced comparator \cite{filippov1991sensitivity}, and the pulse trains are briefly converted into magnetic flux within neurons before undergoing a threshold to produce an output pulse train. The resulting network uses established superconductive fabrication processes \cite{tolpygo2019advanced} and is tuned by an external current, facilitating large scale integration.  All of the circuits presented here are simulated in WRspice \cite{whiteley_wrspice} based on the state-of-the-art MIT LL SFQ5ee fabrication process \cite{tolpygo2019advanced}.

\subsection{Data encoding with SFQ}
\label{encoding}

Because of the phenomenological similarity between neuronal spikes and SFQ pulses, information in the network is represented in the timing and frequencies of SFQ pulses.  Upon sufficiently stimulating a JJ with voltage and/or bias current, a $2\pi$ phase shift is created around a superconducting loop, producing a corresponding SFQ voltage pulse across the JJ, which can be propagated to stimulate downstream JJs.  As a single flux quantum is the smallest possible non-zero voltage pulse, fluxons are ideal information carriers for SNNs, enabling extremely low power neuromorphic processing.  Furthermore, information in a sequence of fluxons is not encoded in the magnitude of the spike but rather in the spike timing as is the case in SNNs and, more notably, the brain.  Fluxons have non-volatile attributes as well, enabling short-term memory and the construction of low power LIF neurons.

A wide range of SNN topologies are available, all of which are amenable to SFQ implementations.  Among these topologies are perceptron networks and feed-forward deep neural networks; both of which can employ stationary input spike rates for classification tasks.  Recurrent neural networks may be constructed as well, enabled by internal time delays in fluxon propagation.  Information coding in recurrent neuromorphic networks is dependent upon the relative timing between spiking events.  Spike-rate-coded networks -- with information encoded in the mean spike frequency -- may be trained with backpropagation similar to a level-based ANN, as described in Section \ref{sec:src_training}.  Online learning using spike timing -- such as spike-time-dependent plasticity \cite{Schuman2022} -- is also amenable to SFQ networks \cite{SegallSTDPSynapse_2023} and may similarly ease training costs.

\subsection{Stochastic-Pass Synapses}

Inspired by the stochasticity in biological systems, the stochastic-pass synapse modulates the connection between upstream and downstream neurons by stochastically gating fluxons.  Whereas conventional synapse approaches modulate the effective amplitude of spiking events, due to the quantized nature of fluxons, it is difficult to accomplish this without conversion between the fluxons and analog current, thereby decreasing system efficiency.  Inspired by the pervasive stochasticity in biological systems \cite{MaassSpikeOrNot, BraunStochasticity}, the proposed synapse stochastically gates incoming spikes, encoding the synaptic weight in the probability that an incoming fluxon will propagate through the synapse.

\begin{figure}[h]
     \centering
     \begin{subfigure}[b]{\linewidth}
         \centering
         \includegraphics[width=0.55\linewidth]{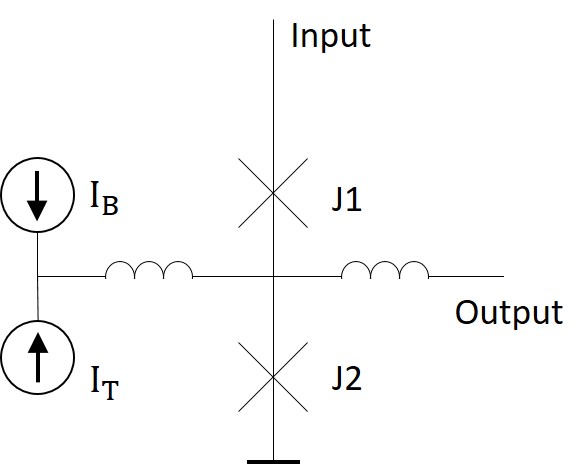}
         \caption{}
         \label{fig:syn_schematic}
     \end{subfigure}
     \\
     \hfill
     \\
     \begin{subfigure}[b]{\linewidth}
         \centering
         \includegraphics[width=\linewidth]{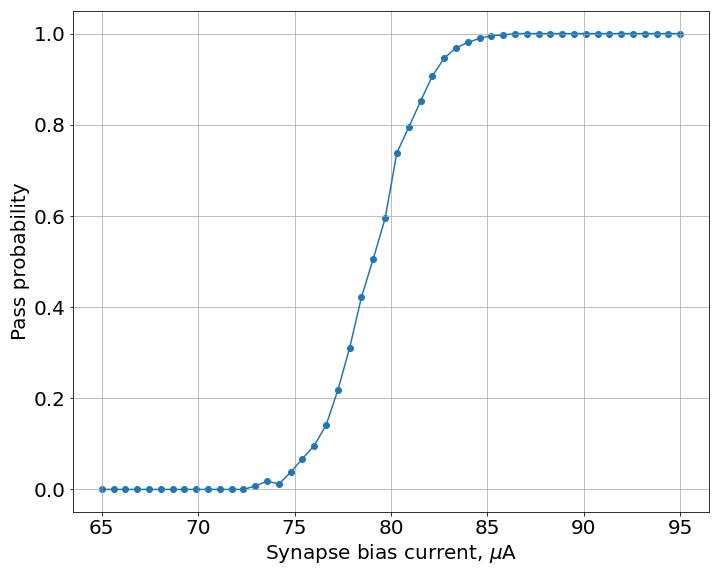}
         \caption{}
         \label{fig:syn_plot}
     \end{subfigure}
        \caption{Synaptic circuit, a) schematic, and b) graph of pass-probability with bias current. The data are presented for $I_C(J_1, J_2)=150~\mu A$, input SFQ pulse frequency of 50 GHz, and T = 4.2~K.}
        \label{fig:synapse_circ}
\end{figure}

The proposed synapse is constructed from a Josephson balanced comparator \cite{filippov1991sensitivity, filippov2021gz}, harnessing thermal noise for true stochasticity.  A Josephson balanced comparator is a pair of serially connected JJs with the bias current, $I_B$, applied between these JJs, as shown in Fig. \ref{fig:syn_schematic}.
When an SFQ pulse is applied to the input, one of the JJs within the comparator undergoes a 2$\pi$ phase shift, depending on the magnitude of $I_B$.  For small $I_B$, $J1$ switches, absorbing the input pulse, whereas for large $I_B$, $J2$ switches, propagating the input pulse to the output.
In the absence of noise current $I_T$, the gating functionality depends deterministically on the applied currents, however in the presence of thermal fluctuations in the applied currents, the balanced comparator exhibits stochastic behavior -- a "grey zone" \cite{filippov1991sensitivity}.
A similar approach has been proposed for SFQ based synapses \cite{semenov2022biosfq}, where a C-SQUID \cite{semenov2003csquid} is used rather than a balanced comparator.

 The synapse weight is encoded in the probability that an input spike will propagate to the output and can be modulated after fabrication by adjusting $I_B$.  Fig. \ref{fig:syn_plot} depicts the probability of passing an incoming SFQ pulse to the output as a function of bias current for specific circuit parameters, demonstrating a continuous swing of weights between $0$ (all fluxons blocked) and $1$ (all fluxons propagated).


\subsection{Leaky Integrate-and-Fire Neurons}

\begin{figure}
    \centering
    \includegraphics[width=1.0\linewidth]{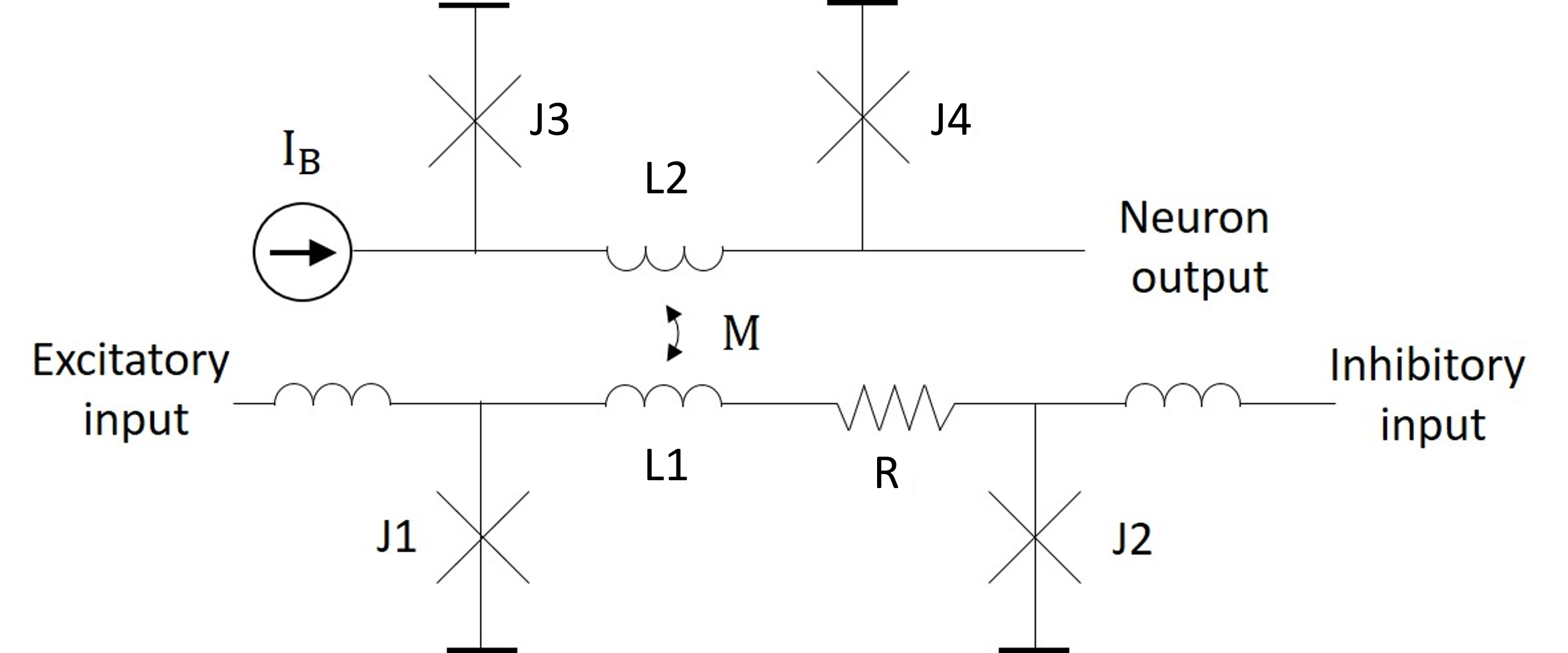}
    \caption{Neuronal circuit.}
    \label{fig:neur_sch}
\end{figure}

\begin{figure}
    \centering
    \includegraphics[width=1\linewidth]{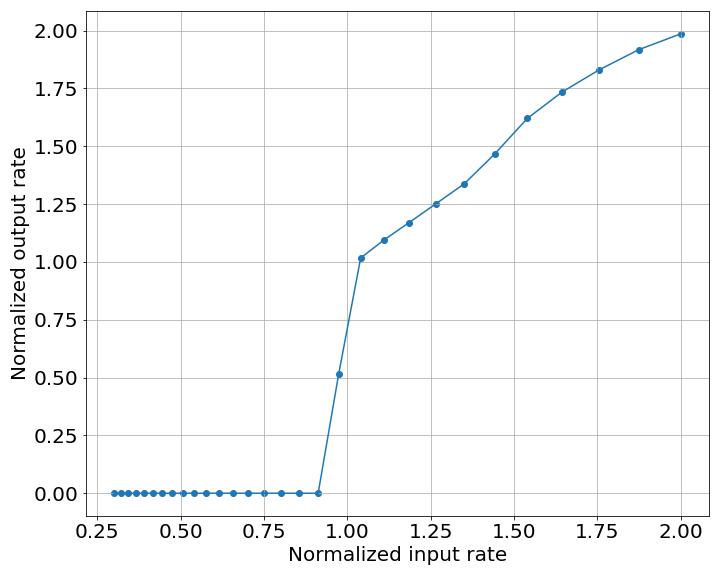}
    \caption{Neuron activation function for asynchronous neuron (see Fig. \ref{fig:neur_sch}). The input and output rates are normalized to approximately 33.3 GHz.}
    \label{fig:neuron_plot}
\end{figure}


Neuronal LIF circuits -- integrating incoming SFQ pulses until an internal threshold is reached and outputting a resultant SFQ spike sequence -- are described here.  A leaky SQUID loop inductively coupled with a firing SQUID loop \cite{krylov2017design} to implement integrating, leaking, and firing behavior.

A conventional SQUID loop with a large inductance is used to integrate incoming SFQ pulses. The input SFQ pulses are applied across $J1$ or $J2$ (indicated in Fig. \ref{fig:neur_sch}) as excitatory or inhibitory inputs respectively.  As $J1$ switches, the fluxon energy is stored as current in inductor $L1$.  This energy is accumulated with additional inputs increasing the flux stored in the loop.  A fluxon applied to the inhibitory input switches $J2$ reduces the flux stored in the loop allowing neurons to have both excitatory and inhibitory interactions.

Leaking may be implemented with the addition of a resistive element $R$ dissipating the energy stored in inductor $L1$.  A resistance on the order of a few ohms produces a linear leakage characteristic, the timing of which can be tuned by changing the size of the resistor. Utilizing the dynamic SFQ (DSFQ) leakage mechanism, additional JJs may be introduced into the loop to provide a faster reset of the circulating current \cite{rylov2019dsfq, krylov2020maj}. 

When the current in the SQUID is sufficiently large, an inductively coupled SQUID produces a firing pulse. A JJ partially biased by an inductively coupled neuron loop current, is shown in Fig. \ref{fig:neur_sch}. Due to the coupling between $L1$ and $L2$, as $L1$ integrates input pulses, current in $L2$ simultaneously increases.  Eventually current in $L2$ -- with contributions from $L1$ and $I_B$ -- is sufficiently large to switch $J4$ producing an output pulse.  In the case where inhibitory inputs are more frequent than excitatory inputs, $J3$ will switch instead of $J4$ sans output spike.  Bias currents through the JJs help to regenerate fluxons, enable signal fan-out, and improve the cascade characteristics.

The relationship between the neuron input and output rates is non-linear, a necessary condition for proper neural network functionality.  Fig. \ref{fig:neuron_plot} depicts the relationship between input and output spike rates (equivalent to the activation function of a conventional level-based neuron if spike-rate-coding is employed). Note the distinct non-linearity akin to a rectified linear unit (RELU), scaled exponential linear unit (SELU), or sigmoidal activation function. The width and slope of the threshold region as well as the saturation characteristics are adjusted by tuning $I_B$, $L1$, $L2$, $R$, and $I_C (J1)$ (see Fig. \ref{fig:neur_sch}).  The threshold input rate may be tuned after fabrication through the application of bias input sequences.  At higher frequencies, output rate saturates as $J4$ moves toward the resistive regime adding additional non-linearity.


The input fan-in is realized using conventional RSFQ confluence buffers (pulse mergers) \cite{likharev1991rsfq}. These buffers exhibit a saturating spike rate as multiple input pulses arriving in close succession produce only one output pulse. This property assists in the saturation of the neuronal output spike rate.

Negative weights are applied as inhibitory inputs to the neuron. A single synapse can therefore be implemented as a differential pair of synapse circuits connected to the excitatory and inhibitory inputs of a neuron.
Inhibitory or excitatory bias input spike sequences may be added to adjust the neuron threshold.

Fan-out is managed by splitter trees, which could incur well-known area and delay overhead \cite{SchneiderFan}. Multiple techniques exist to reduce the overhead of the signal fanout in large scale SFQ circuits \cite{jabbari2021splitter}, \cite{katam2017fanout}. These techniques are primarily based on utilizing a splitter with more than two outputs at the cost of reduced parameter margins.

\subsection{Cascading Synapses and Neurons for Deep Networks}

The SFQ neuromorphic primitives can be cascaded to construct deep neuromorphic networks enabling straightforward circuit design with minimal layer-to-layer tuning. As inputs and outputs from synapses, neurons, fan-in, and fan-out circuits are all SFQ pulses there is no need for costly signal conversion to construct large multilayer networks.

To ensure that spiking activity is similar from layer to layer, spiking activity can be regenerated through the use of signal confluence and additional input spike sequences.  When an RSFQ confluence buffer merges two or more spike sequences, the output spike rate is the combined rate of the input sequences, increasing signal activity.  Furthermore, additional bias spike sequences may be incorporated to further regenerate signal activity.

\begin{figure}
    \centering
    \includegraphics[width=0.5\textwidth]{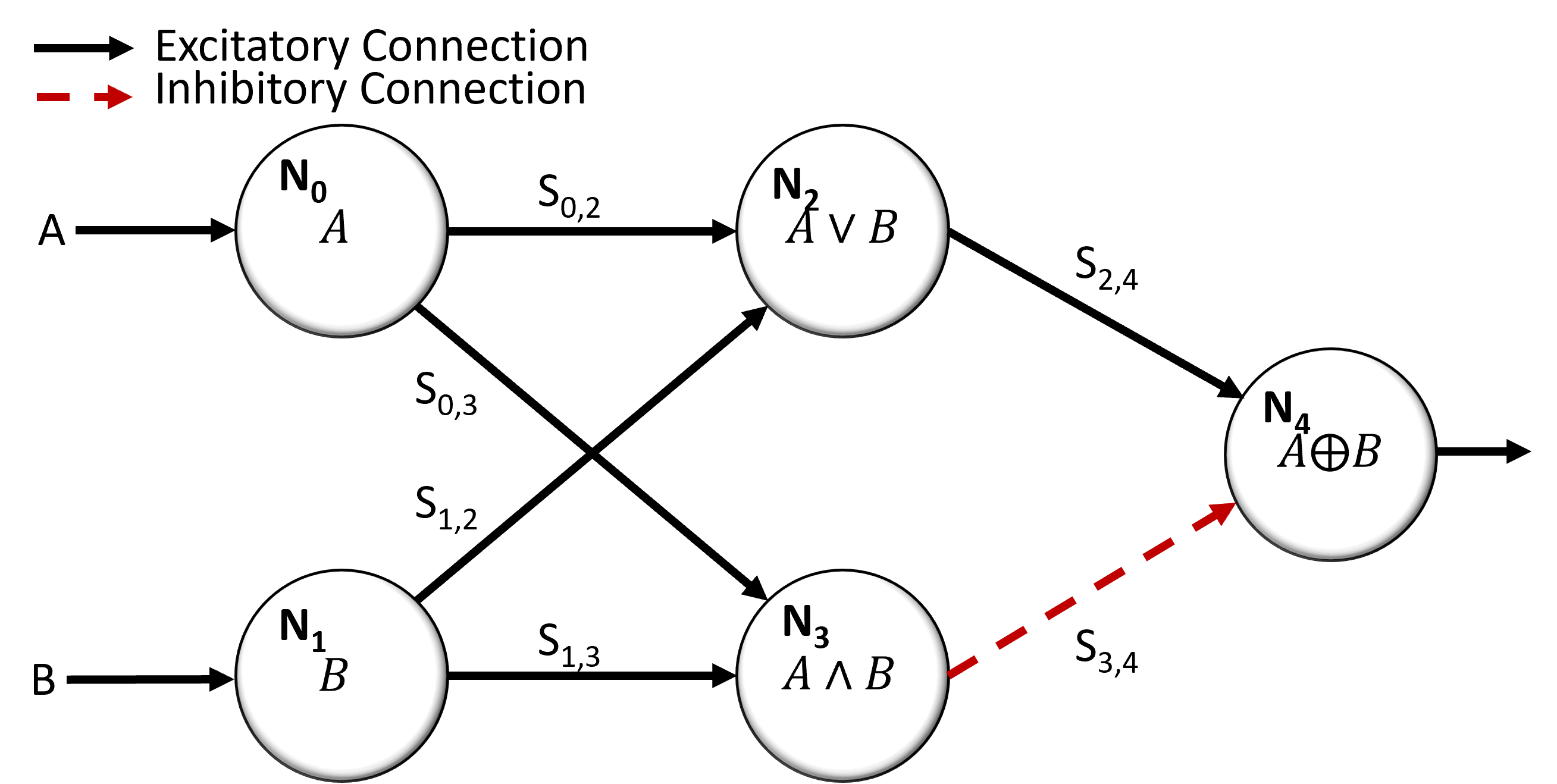}
    \caption{XOR spiking neural network. The primary inputs, bias input, and primary outputs are represented as spike frequencies in proportion to the saturation output neuronal rate.}
    \label{fig:xor}
\end{figure}

\begin{figure*}
    \centering
    \includegraphics[width=0.9\textwidth]{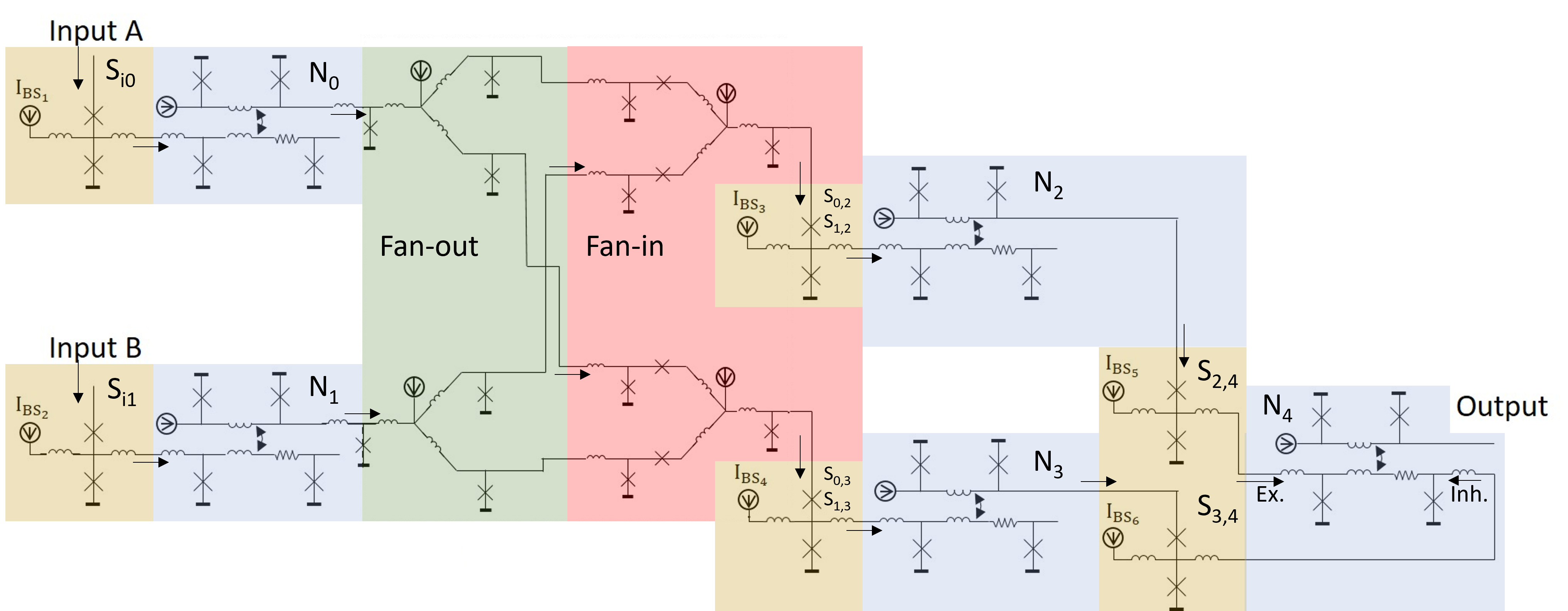}
    \caption{Neuromorphic SFQ network computing a two bit XOR function with synapses in yellow, neurons in blue, fan-out in green, and fan-in in red.}
    \label{fig:xor_schematic}
\end{figure*}

Layer-to-layer spiking activity can be tuned at a network-architecture level, mitigating the need to individually tune circuit parameters for each layer or for a specific network architecture.  Specifically, synapse weights may be tuned to ensure that spiking activity remains within proper regions of operation for the circuits.  Furthermore, the rates of the bias input spike sequences may be tuned through the use of synapse circuits.


\section{Case Study: Demonstration of Two-layer XOR Neural Network}

The first demonstration of a multi-layer fully SFQ neuromorphic network is described here.  The network is trained to compute the XOR functionality, and demonstrates large non-linearity and input separation.

\subsection{Training and Weight Mapping of XOR Network}

\begin{figure*}
    \centering
    \includegraphics[width=0.75\textwidth]{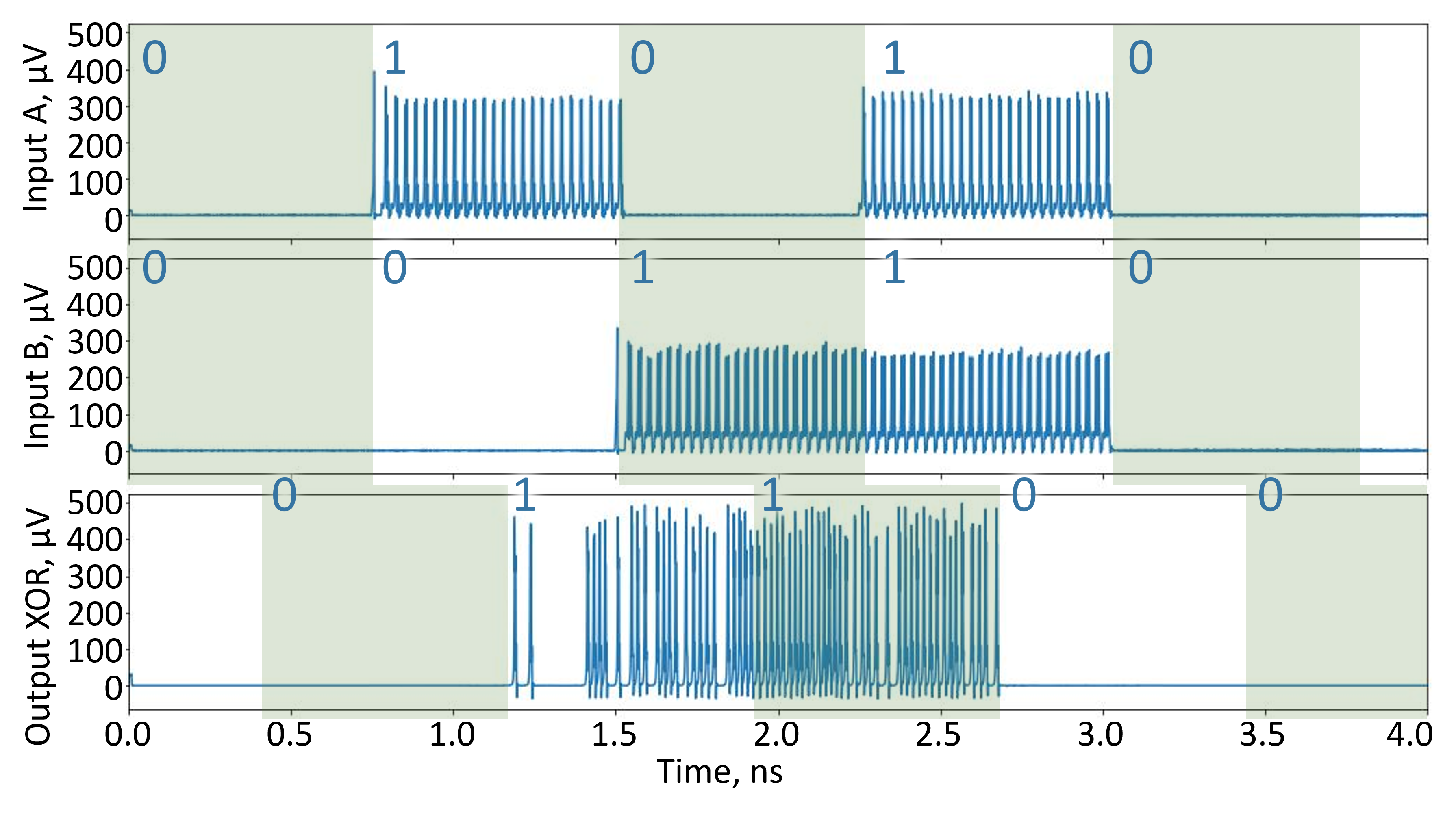}
    \caption{Spiking behavior with different input stimuli for XOR network.}
    \label{fig:xor_operation}
\end{figure*}

A trained neuromorphic network to compute the two bit XOR function requires at least two layers in an ANN along with negative weights and bias inputs.  The chosen network is depicted in Fig. \ref{fig:xor}.  Input neurons 0 and 1 normalize the network inputs rates to the neuronal output levels.  Due to trained neuronal biases, neuron 2 will have a high output when one of the two inputs is on while neuron 3 requires both inputs to be on in order to activate.  Neuron 4 uses inhibitory weights to compute the XOR functionality, activating only if neuron 2 is active while neuron 3 is not.  Neuron 4 will therefore only turn on when exactly one of the inputs is on: the XOR functionality.

Spike-rate-coding was chosen to map the SFQ neuromorphic primitives to the chosen network.
As described in Sections \ref{sec:src_training} and \ref{encoding}, spike rate coding supports training similar to a level-based ANN and allows direct mapping of architecture and trained weights between the level-based ANN and the spike-rate-coded SNN.  The two network inputs are stationary pulse trains with rates close to neuronal saturating rates, encoding a logic $0$ ($1$) as a low (high) spike rate. The synapse weights are encoded in the probability of the spike propagation. The bias input sequences are trained to make the neurons sensitive to different combinations of inputs, tuning the threshold input rate of Fig. \ref{fig:neuron_plot} to ensure large separability between the input patterns.   The network outputs may be interpreted by the average spike rate in proportion to the saturating spike rate of the neurons.

\subsection{Circuit topology of multilayer network}
The XOR network was directly implemented with the SFQ primitives described in Section \ref{sec:3}.
A schematic of the neuromorphic network based on these components, computing a two bit XOR function (presented in Fig. \ref{fig:xor}) is shown in Fig. \ref{fig:xor_schematic}.
The synaptic weights and neuronal biases are tuned by changing the bias current.



\subsection{Network Results}

\begin{figure}[H]
    \centering
    \includegraphics[width=1.0\linewidth]{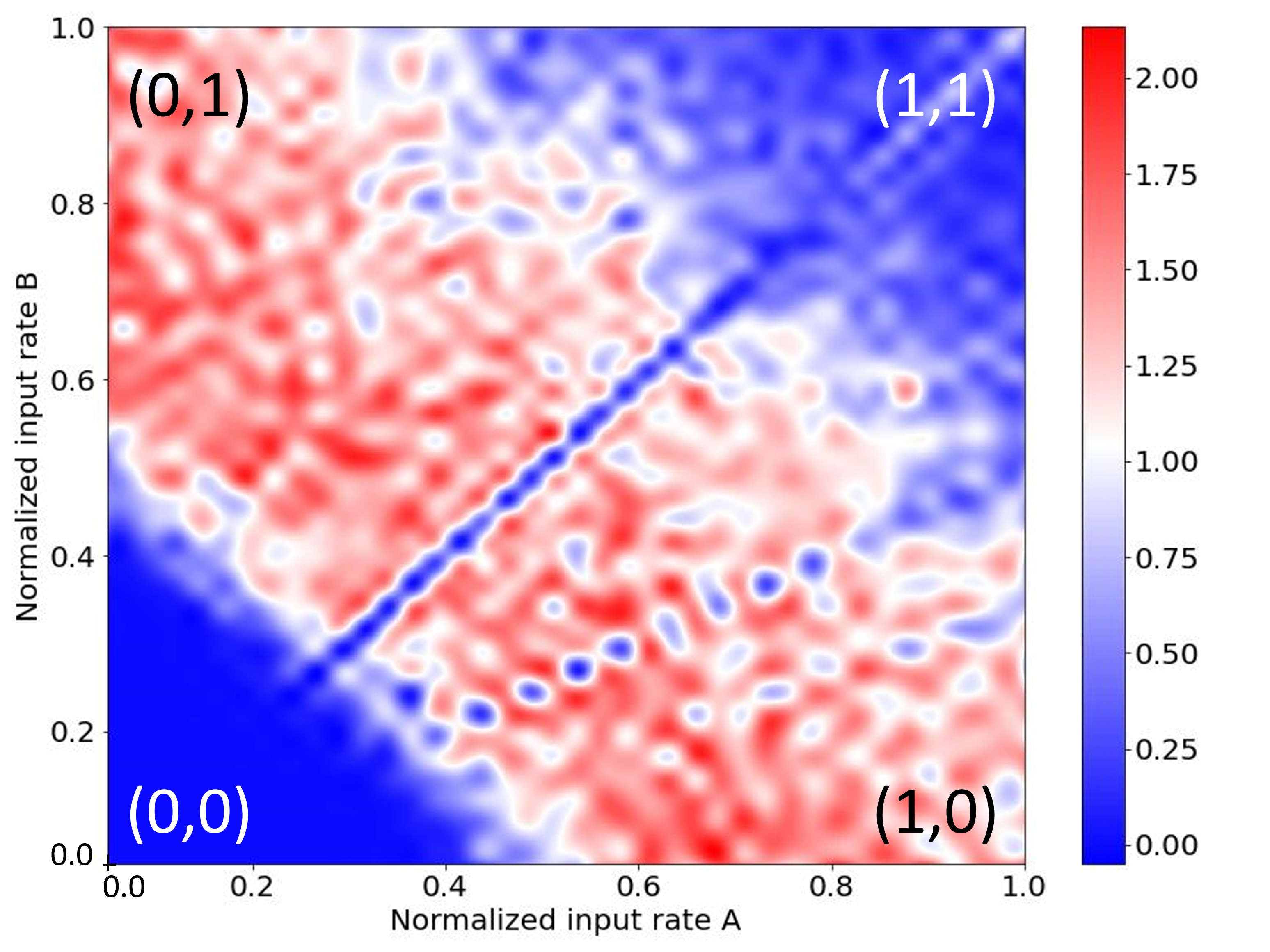}
    \caption{Phase diagram of classification for XOR network output.  Red represents high output rates (between 1 and 2), blue represents output rate between 0 and 1 with the colors in between representing the domain boundaries. The input and output rates are normalized to a pulse rate of approximately 33.3 GHz.}
    \label{fig:phase}
\end{figure}

As shown in Fig. \ref{fig:xor_operation}, the network correctly computes the XOR function.  Additionally the network is robust to variations in the input rates, showing a large separation between classes.  A phase diagram of classification for the XOR network is shown in Fig. \ref{fig:phase}a, where the dependence of the network output rate is shown as a function of the two input rates. It is desirable for the high output rates corresponding to the output of logic $1$ (shown in Fig. \ref{fig:phase} in red) to map to the input rates corresponding to logic $10$ and $01$. The input rates corresponding to logic $00$ and $11$ should produce low spike activity (shown in blue).  This classification diagram displays good separation of the output states with respect to the input spike rates.
The proposed network is therefore robust to variations in the input rates.


\section{Conclusions}

The first demonstration of a fully SFQ multi-layer neuromorphic network is presented here along with SFQ neuromorphic primitive circuits that can be directly cascaded to construct a broad range of network architectures.  While the demonstration shows one feed-forward network using spike rate coding, the circuits are amenable to richer representations of information in spiking neuromorphic networks including recurrent neural networks, spike-time-dependent networks, and online learning.  As network layers can regenerate spiking behavior, deep network architectures are readily attainable through natural cascading of successive layers.

The proposed network -- entirely based in available SFQ technologies -- has the advantage of being extremely energy efficient as compared with conventional CMOS technologies \cite{holmes2013energy}, \cite{bozbey2020single}.  Furthermore, as conversions between fluxons and analog currents are constrained within each individual neuron, the network is compact and scalable. Additionally, the proposed scheme does not require unconventional devices or complex 2.5-D or 3-D integration, and can be produced using standard niobium fabrication processes.


\bibliographystyle{IEEEtran}
\bibliography{IEEEabrv,main}

\end{document}